\documentclass[aps,preprint,showpacs,groupedaddress]{revtex4}  
\usepackage{graphicx}  

\usepackage{amssymb}   

\begin{document}

\title{A new approach to test Lorentz invariance}
\author{H.W.~Wilschut}
\author{E.A.~Dijck}
\author{S. Hoekstra}
\author{K. Jungmann}
\author{S.E.~M\"uller}
\author{J.P.~Noordmans}
\author{C.J.G.~Onderwater}
\author{C.~Pijpker}
\author{A.~Sytema}
\author{R.G.E.~Timmermans}
\author{K.K.~Vos}
\author{L. Willmann}
\address{Kernfysisch Versneller Instituut, University of Groningen,
 Zernikelaan 25, 9747 AA Groningen, The Netherlands}
\email{Wilschut@kvi.nl}
\date{\today}


\begin{abstract}
Lorentz invariance  in the weak interaction has been tested rather
poorly compared to the electromagnetic interaction. In this work
we show  which tests on the weak interaction should be considered.
We focus on one particular test that explores the spin degree of
freedom in $\beta$ decay. To connect various phenomenological
tests of Lorentz invariance in the weak interaction, we exploit a
new theoretical model that may provide a framework that relates the
different tests.
\end{abstract}

\pacs{11.30.Cp, 24.80.+y, 23.40.Bw}

\maketitle

There are many experimental tests of Lorentz
invariance~\cite{Kostelecky:2008ts}.
Arguably, the most precise tests have been done for the electromagnetic interaction.
For the weak interaction, in particular for $\beta$ decay, very
few studies have been made. This despite the fact that the
Standard Model~(SM) originated from --and has been shaped by-- the
details of the weak interaction, i.e. the violation of Parity (P)
and Charge  conjugation (C) on one hand and the violation of the
combined CP symmetry on the other.

Presently, one of the main efforts in fundamental physics is the
unification of the Standard Model with General Relativity, in what
is mostly referred to as quantum gravity models. Certain models of
quantum gravity contain terms which violate Lorentz invariance and
CPT symmetry (e.g.~\cite{Kostelecky:1988zi,Kostelecky:1995qk,Ellis99,Burgess02,Gambini99}).
Manifestations of Lorentz Invariance Violation (LIV) could be
searched for in low-energy  experiments, such as in $\beta$ decay. Requiring a theory that identifies the appropriate observables.
Kosteleck\'y and coworkers have developed a theoretical framework
named ``Standard-Model Extension'' (SME) that contains all the
properties of the Standard Model and General Relativity, but
additionally contains all possible terms violating Lorentz and CPT
symmetry via spontaneous breaking of Lorentz
invariance~\cite{Colladay:1998fq}. It follows from this phenomenological
approach, that observables for the different interactions are a priori independent. In this respect it is insufficient to test
only the electromagnetic interaction.

We have started an experimental and theoretical program on LIV
considering charged currents in the weak interaction, focusing on
$\beta$~decay. Recently a theoretical framework has been
formulated that gives guidance to possible experiments~\cite{JN2013}. It also
shows to what extent various experiments could be related. In this
theory the Lorentz symmetry breaking is implemented by modifying
the propagation of the W boson. The theoretical
motivation for this can be found in reference~\cite{JN2013}. Here we will
discuss the relevant results for $\beta$-decay experiments. In our
experimental work we focus on the spin degree of freedom which up
to now was not  considered at all.

In the SM the $\beta$-decay rate, ignoring Coulomb and induced recoil
effects,  is given by~\cite{Jackson}
\begin{equation}\label{SMbeta}
    \frac{d\Gamma}{\Gamma_0}= 1 + \vec{\beta}\cdot[A\frac{\langle\vec{J}\rangle}{J}+G\vec{\sigma}],
\end{equation}
where $\vec{\beta}$ is the velocity of the $\beta$~particle in
units of the light velocity. $\frac{\langle\vec{J}\rangle}{J}$ is
the degree of nuclear polarization of the parent nucleus and its
direction. $\vec{\sigma}$ is the spin vector of the
$\beta$~particle. \ $\hbar/\Gamma_0$ is the lifetime of the
nucleus. $A$ and $G$ are the well known parity violating
parameters: $A$ is the $\beta$ asymmetry or ``Wu'' parameter and
$G$ the longitudinal polarization of the outgoing $\beta$~particle
($G=\pm 1$ for $\beta^{\mp}$).

If there is a preferred direction in space (i.e.
Lorentz symmetry breaking), eq.~(\ref{SMbeta}) will be modified. In that case we
expect it to be of the form
\begin{equation}\label{LIVbeta}
    \frac{d\Gamma}{\Gamma_0}= 1 + \vec{\beta}\cdot[A\frac{\langle\vec{J}\rangle}{J}+ \xi_1 \hat{n}_1 +G\vec{\sigma}] + \xi_2 \frac{\langle\vec{J}\rangle}{J}\cdot\hat{n}_2 +\xi_3\vec{\sigma}\cdot\hat{n}_3.
\end{equation}
Here $\hat{n}_i$ are the preferred directions in space.  The
directions do not need to be the same for the various observables,
as we will show below. The $\xi_i$ are  the magnitudes of the
Lorentz symmetry breaking terms. Because a measurement of
$\vec{\sigma}$ inevitably involves measuring $\vec{\beta}$, we
only consider $\xi_1$ and $\xi_2$. Where $\xi_1$  measures the degree of
$\beta$-emission anisotropy of non-oriented nuclei, while $\xi_2$
measures the dependence of the lifetime on orientation. With respect to eq.~\ref{LIVbeta} two important observations should be made. First, because of the low velocities of the parent nucleus we do not consider modifications of the decay rate due to
boosts i.e. a dependence on absolute velocity. Second, the dependencies predicted in \cite{JN2013} go beyond those given in this equation, however, for the experiments we discuss here it suffices.

Concerning the $\beta$ asymmetry, $\xi_1$, two measurements were
made in the seventies~\cite{Newman:1976sw,Ullman:1978xy}. Both
were made  for forbidden decays. The main idea behind these
experiments was to test  rotational invariance by trying to
observe violation of angular momentum conservation. A forbidden
decay would become less forbidden to the extent that
angular momentum would not be conserved, thus relatively enhancing
the violating signal. By measuring the decay rate in various
directions and correlating it with the earth's rotation,
deviations from isotropy were searched for. To reach high
precision, the whole setup including source and detector needed to
be rotated. No deviations were found with a dependence of
$\cos(\omega t)$ up to a level of $1.6\times 10^{-7}$  for the
unique first forbidden transition in $^{90}$Y
\cite{Newman:1976sw}, where $\omega$ is the earth's rotation
frequency. The second experiment considered a second forbidden
transition in $^{99}$Tc \cite{Ullman:1978xy} and reached a limit
of $3\times 10^{-5}$. It remains to be seen what these values mean
in an underlying theory. It is presently evaluated \cite{Jacob}
within the context of our theoretical work \cite{JN2013}. Some
preliminary conclusions will be discussed below.

\begin{figure}
  \centering\includegraphics[width=0.8\columnwidth]{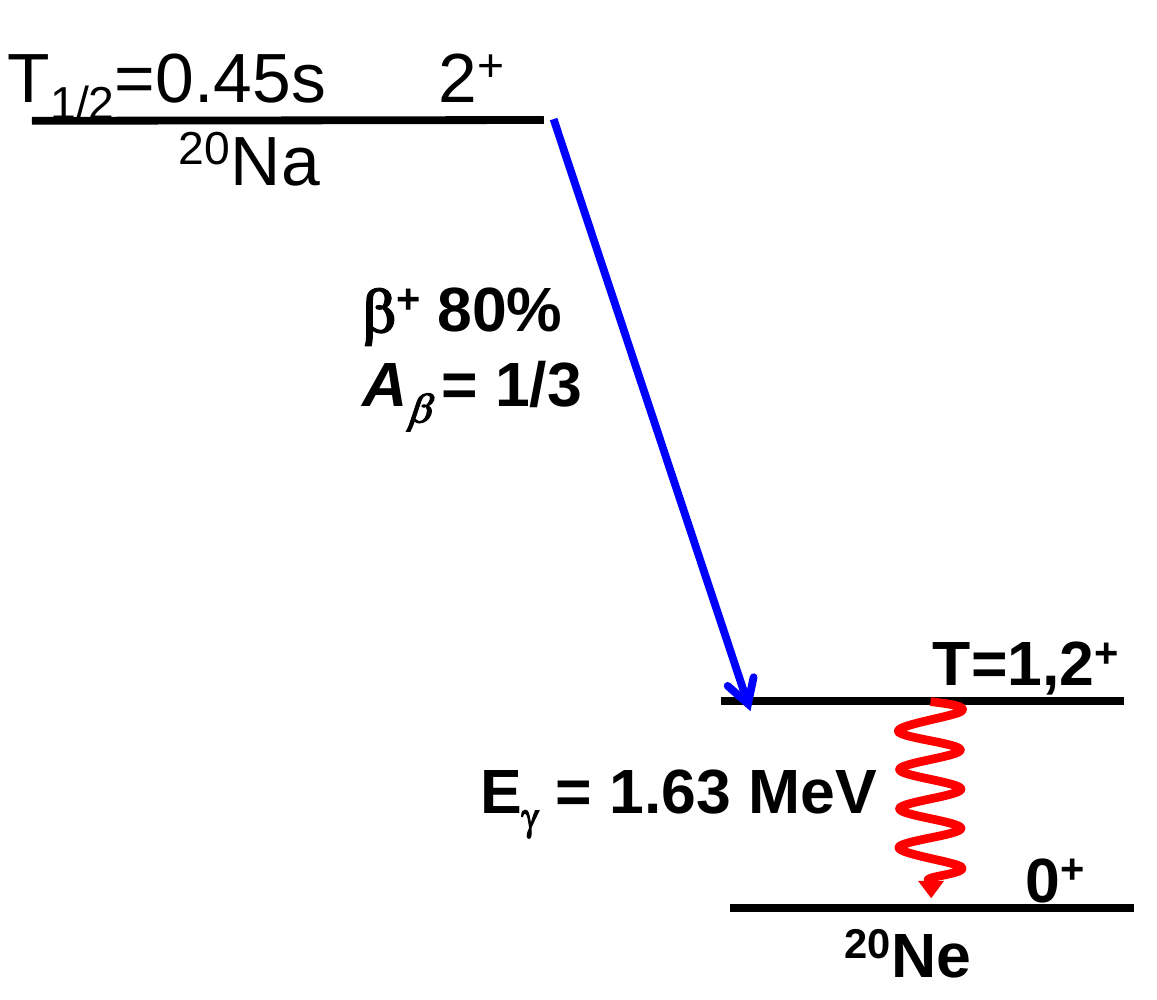}\\
  \caption{Relevant part of the decay scheme of $^{20}$Na}\label{levelscheme}
\end{figure}

In our experimental work we are considering a polarization-dependent
lifetime, parameterized by $\xi_2\hat{n}_2$, for which no
experimental information is available as yet \cite{neutron}. It requires a sample
of radioactive nuclei with oriented spin,
for which the lifetime must be measured. We have searched for
methods where these two tasks could be efficiently done. We found
a class of nuclei that allows one to measure the polarization
independently from the lifetime. The nuclear polarization is best
measured from the asymmetry parameter $A$. The lifetime can be
obtained from measuring the depopulation of a nuclear excited
state that is fed by the $\beta$~decay. We show in
fig.~\ref{levelscheme} schematically the situation for $^{20}$Na
as an example.
We define $P\vec{j}=\frac{\langle\vec{J}\rangle}{J}$, where
$\vec{j}$ is a unit vector in the direction of $\vec{J}$ and $P$
the degree of polarization. Building an asymmetry by measuring the
difference of decay rates $R_\beta$ between the emission of
$\beta$ particles parallel and antiparallel to $\vec{J}$ with an
analyzing power $K$, one determines $P$ by
\begin{equation}\label{PA}
    P=\frac{1}{AK}\frac{R_\beta^\uparrow - R_\beta^\downarrow}{R_\beta^\uparrow + R_\beta^\downarrow},
\end{equation}
which allows one to extract $\xi_2$ by measuring the
$\gamma$-decay rates for the two polarization directions
\begin{equation}\label{xi2}
\xi_2(\vec{j}\cdot\hat{n}_2)=\frac{1}{P}\frac{R_\gamma^\uparrow - R_\gamma^\downarrow}{R_\gamma^\uparrow + R_\gamma^\downarrow}=\frac{1}{P}\frac{\tau^\downarrow - \tau^\uparrow}{\tau^\uparrow + \tau^\downarrow}.
\end{equation}
To be independent of possible changes in the sample sizes for the
two spin directions, one  may choose to fit the lifetimes of the
sample instead, as noted in the last equality. We  assume  that
the electromagnetic and strong interaction are not breaking
Lorentz symmetry. The yield of photons may depend on the degree of
polarization but not on its sign.
\begin{figure}
  \centering\includegraphics[width=0.8\columnwidth]{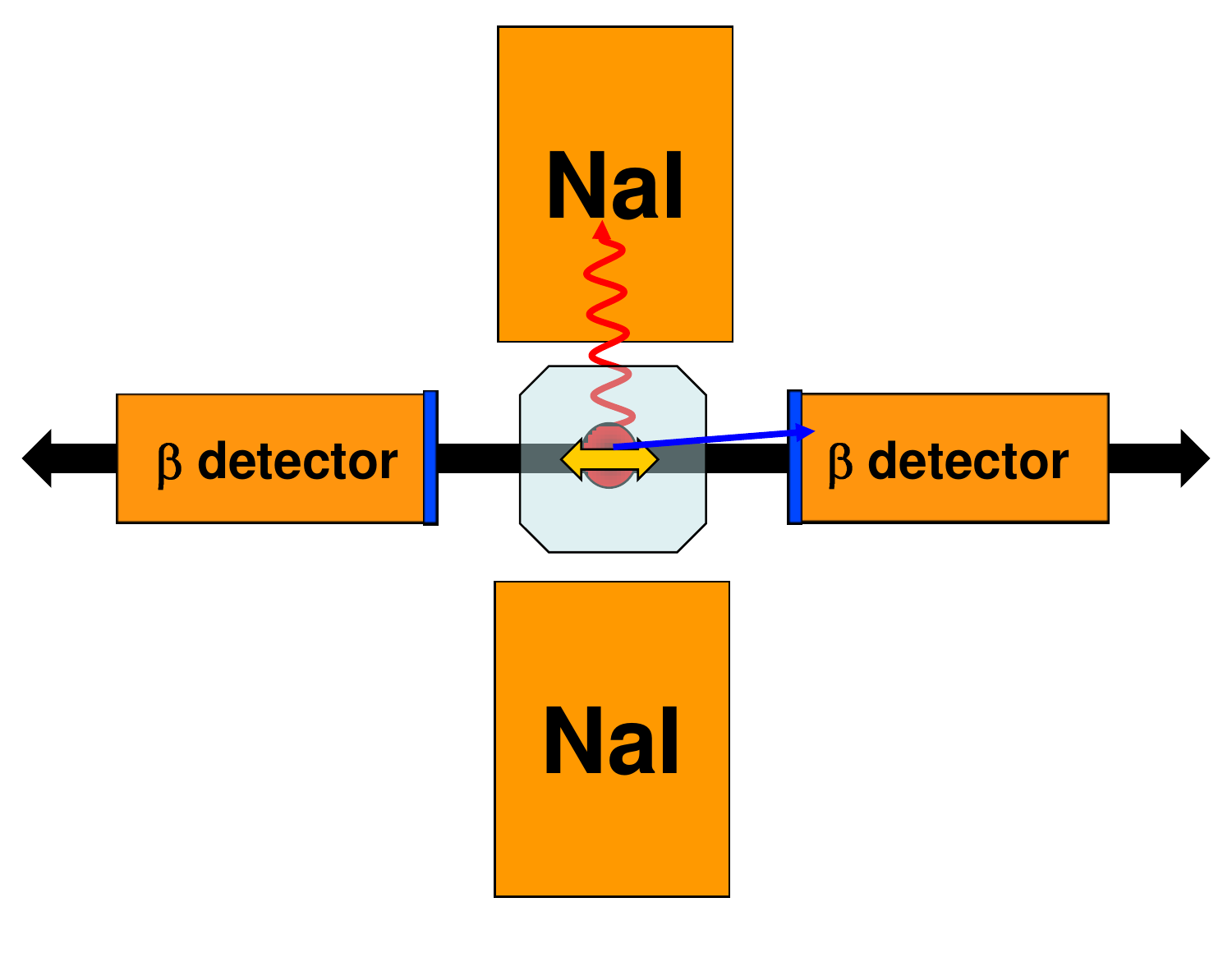}\\
  \caption{Generic setup to measure rotational
  invariance violation in spin-polarized nuclei.
  The yellow arrow indicates the polarization directions of the radioactive sample.
  The complete setup needs to be oriented  in the laboratory, symbolically indicated by the black arrow,
  to probe specific ``preferred'' directions as discussed with reference to fig.~\ref{oscillating}.}
  \label{setup}
\end{figure}
 Detecting $\gamma$'s has, therefore, the advantage that
 systematic errors due to parity violation are eliminated,
while evaluating eq.~(\ref{xi2}) from the $\beta$-particle yield
might lead to errors mimicking LIV.

 A generic setup is
shown in fig.~\ref{setup} which measures simultaneously the
asymmetry in $\beta$ and $\gamma$ emission from a sample where the
polarization can be efficiently reversed.  Many of the systematic
errors can be eliminated with high precision in this highly
symmetric setup  cf.~\cite{Brantjes}. Although there are many
isotopes that can be studied with the strategy described above,
the requirement of polarization and the demand of a high source
strength restricts the isotope choice.

\begin{figure}[t]
  \includegraphics[width=\columnwidth]{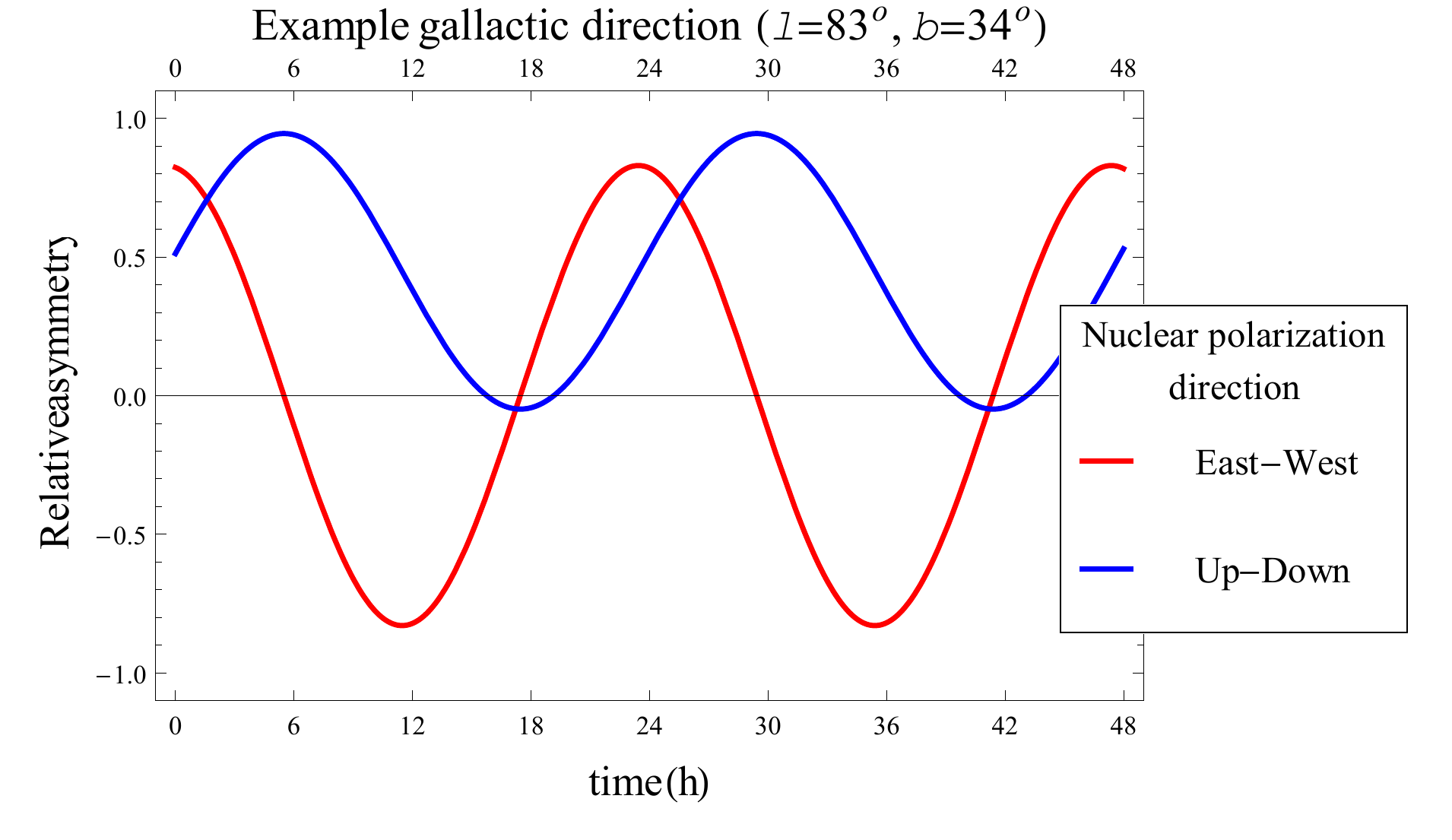}\\
  \caption{Example of a lifetime asymmetry as function of time for two polarization directions
  with an arbitrary chosen ``preferred direction'' as indicated}\label{oscillating}
\end{figure}

The direction of polarization, $\vec{j}$, should be chosen in the
context of systematic errors. For example, choosing the
polarization parallel to the earth's rotation axis would result in
an asymmetry of the $\gamma$ yield independent of the time
of day. In contrast, orienting $\vec{j}$ in the east-west
direction means the asymmetry would show as a sinusoidal
dependence around zero with the pattern reversing sign every half
rotation of the earth. Two cases are shown in
fig.~\ref{oscillating} for a hypothetical ``preferred direction'';
these are polarizations in the east and west directions and
polarization in the up and down directions, i.e. perpendicular to
the earth's surface. In the latter case a signal would have a zero
offset and the sinusoidal dependence would be shifted with respect
to the east-west case.
It appears that the east-west configuration has to be preferred
with respect to systematic errors, because a constant offset is
difficult to discriminate from a systematic bias. However, in this
east-west configuration, no signal would be observed if
$\hat{n}_2$ is parallel to the earth's rotation axis. Therefore,
more than one orientation of the setup of fig.~\ref{setup} should
be considered.

Measuring $\xi_2$ requires a highly active sample that can be
polarized. Our laboratory has an intense solid-state laser for
efficient trapping of radioactive $^{21}$Na in a magneto-optical
trap \cite{wilschut}. This laser can also be used for polarizing
large samples of any Na isotope. Polarizing the sample in a buffer
gas avoids  the loss mechanisms associated with capturing ions,
and neutralizing them for atomic trapping, hence the buffer-gas
method is preferred. The decay scheme shown in
fig.~\ref{levelscheme} is available in $^{20,24-27}$Na. Of these
$^{20,26,27}$Na are useful in our setup (see below), but only
$^{20}$Na can be produced in excess of $10^6$ particles per second
at our facility, for this reason $^{20}$Na was selected for this
study.

\begin{figure}[bh!]
  \includegraphics[width=\columnwidth]{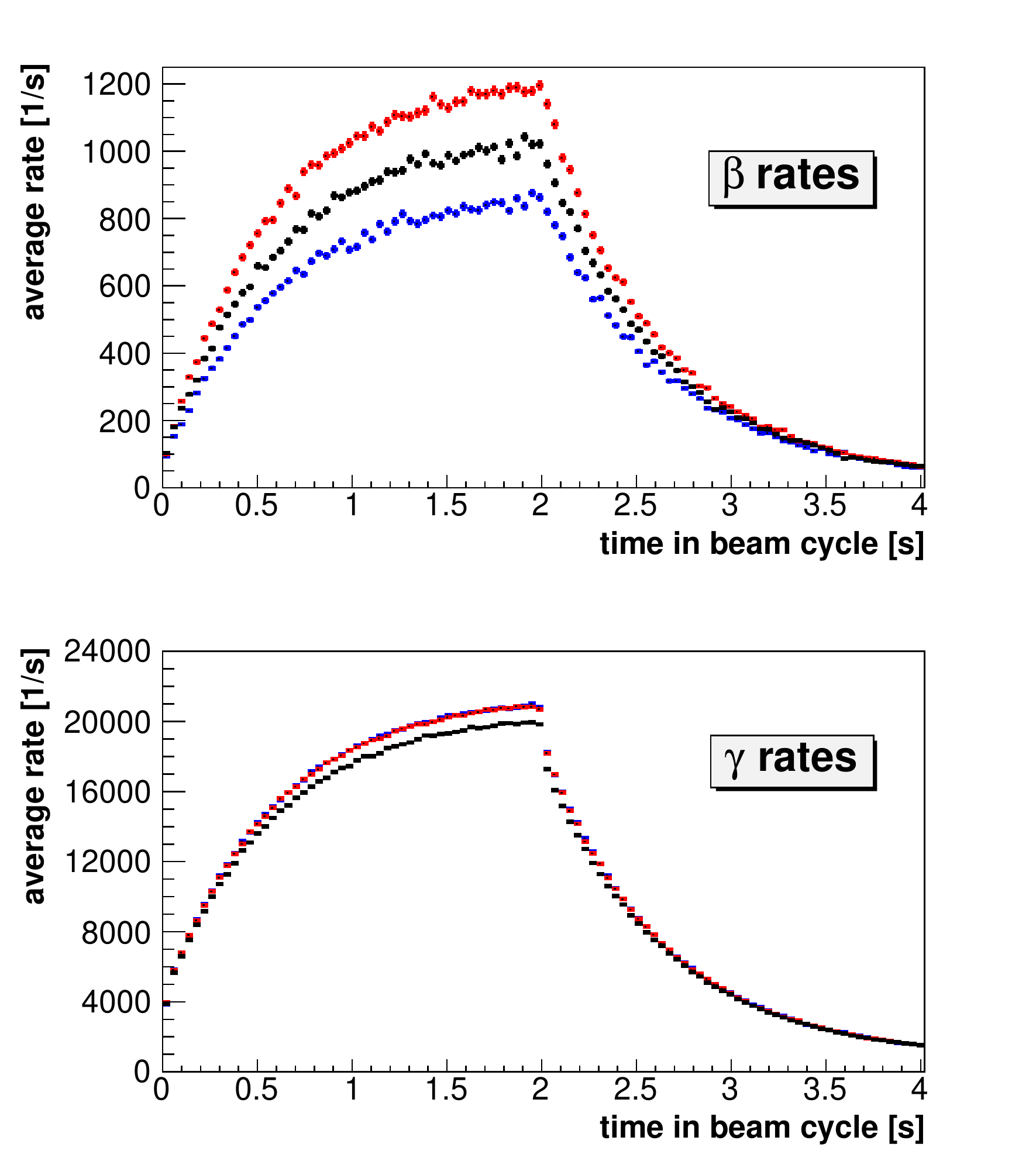}
  \caption{$\beta$ (top) and $\gamma$ (bottom) rates measured
  with laser polarization in two directions (indicated by red and
  blue data points, the latter mostly covered by the former)
and without laser (black data points). The periods with beam  on
and off are 2 seconds.} \label{Aukeinput}
\end{figure}

In this  experiment we look for a change in the decay rate of the
allowed $\beta$ decay of $^{20}$Na when reversing the orientation
of the nuclear spin $\vec{J}$ via optical pumping. ${}^{20}$Na is
produced via the ${}^{20}$Ne(p,n)${}^{20}$Na reaction by colliding
a ${}^{20}$Ne beam with hydrogen in a gas target \cite{Traykov}.
The resulting isotopes pass through the TRI$\mu$P isotope
separator facility to obtain a ${}^{20}$Na beam which is stopped
in a buffer gas cell filled with up to 8 atm of neon gas.

Adjustable aluminum degrader foils in the beam line allow to
position the beam's stopping distribution in the center of the gas
cell. The neon buffer gas is cleaned with a cryo-trap filled with
liquid nitrogen and a gas purifier cartridge. A heatable dispenser
with natural sodium is mounted inside the buffer gas cell. The use
of the dispenser proved to be essential. The natural sodium binds
residual chemically active contaminants in the gas  that would
bind radioactive sodium into molecules, making them unavailable
for polarization.

The stopped ${}^{20}$Na atoms in the center of the buffer gas cell
are optically pumped into a ``stretched'' state in which the
electronic and nuclear spins are both aligned along the direction
of the magnetic holding field provided by Helmholtz coils. To
achieve this, a circularly polarized laser beam with \mbox{589 nm}
wavelength is sent through the buffer gas cell. Remotely operated
beam blockers allow to switch the polarization of the laser light
going to the cell. Depending on the helicity of the light that
enters the buffer gas cell, the atoms will be pumped into a state
with the spins aligned or anti-aligned to the direction of the
magnetic field. The polarization is measured from $\beta^{+}$
rates as shown in fig.~\ref{Aukeinput} using eq.~(\ref{PA}). For
better control of systematic errors we operate the setup in three
short cycles of 4 seconds, one for each spin polarization
direction and one with the laser beam off.
 Within each cycle the beam is on and off for
two seconds. A maximum polarization of about 50\% was obtained.
The polarization decreases when the beam is off, presumably due to
the drift of particles out of the volume covered by the laser. For
this reason a short lifetime is advantageous; The halflife of
$^{20}$Na is 448 ms. These and other factors playing a role in the
experimental method will not be discussed further here but in a
forthcoming publication \cite{Mueller}.

\newpage
Fig.~\ref{Aukeinput} shows also the measured $\gamma$ rates. When
the sample is polarized, a small enhancement of the $\gamma$
emission can be observed, which is independent of the sign of the
polarization. This is due to the quadrupole  emission pattern of
the $\gamma$-rays following the $\beta$ decay of the polarized
parent nucleus. Of course this signal may not depend on the sign
of the polarization. A first set of data for up-down polarization
has been analyzed and the results will be available
soon~\cite{Mueller}. In this data set the precision is at a
level of $10^{-3}$.

In the following we give some details of the theoretical work that
we use to give more physical meaning to the experimental
observables. Indeed, to have limits on $\xi_i \hat{n}_i$ without
being able to relate them to each other or to some underlying
theory is not satisfactory. For this reason a version of the SME
was formulated applicable for weak decays. We use an extension where only the
W propagator is modified. Such a specific choice, of course, can
not cover all theoretical possibilities, but can be quite well
motivated and is most appropriate for $\beta$ decay. This
theoretical work is discussed in ref.~\cite{JN2013}. Here, we
restrict the discussion to the results of this theoretical
exploration, in particular for the parameters $\xi_i\hat{n}_i$.

The Lorentz-violating propagator at low energies
that we use is given by
\begin{equation}
\left\langle W^{\mu+}(p)W^{\nu-}(-p)\right\rangle = \frac{-i(g^{\mu\nu}+\chi^{\mu\nu})}{M_W^2} \ ,
\label{wpropagator}
\end{equation}
where $g^{\mu\nu}$ is the Minkowski metric and $\chi^{\mu\nu}$ is
a general Lorentz-violating (complex, possibly momentum-dependent)
tensor. Neglecting the dependence on boosts, i.e. assuming that
the velocities of the parent nuclei are small with respect to the
``preferred frame'', we find that the parameter for the anisotropy
of the emission direction is given by
\begin{equation}\label{xiandchiF}
    \xi_1\hat{n}_1^l = 2\chi_r^{0l}
\end{equation}
for Fermi transitions, while for Gamow-Teller transitions
\begin{equation}\label{xiandchiGT}
    \xi_1\hat{n}_1^l = \frac{2}{3}(\chi_r^{l0}+\epsilon^{lmk}\chi_i^{mk}).
\end{equation}
Here the subscript $r$ ($i$) refers to the real(imaginary) part of
$\chi$; the notation is such that, for example, the right-hand
side of eq.~\ref{xiandchiGT} has the $x$ component
$\frac{2}{3}(\chi_r^{10} + \chi_i^{23}-\chi_i^{32})$.

Thus, a striking result of this theory is that there is no single
``preferred direction''. It differs for Fermi and Gamow-Teller
transitions, but it is also different for other observables. How
$\xi_1\hat{n}_1$ depends on $\chi^{\mu\nu}$ for forbidden $\beta$
decays, is presently evaluated. An enhancement for a nucleus with
atomic number $Z$ and radius $R$ can be expected of order $\alpha
Z/R \approx 0.3Z$, i.e. an order of magnitude larger than the LIV
effect in allowed decays \cite{Jacob}. The underlying idea of
ref.~\cite{Newman:1976sw,Ullman:1978xy} that angular momentum may
not be conserved in the weak interaction can thus be made
quantitative in the present theory. The theory also shows that
searching for LIV in second-forbidden reactions as in
\cite{Ullman:1978xy} gives no further enhancement over
first-forbidden decays.

Gamow-Teller transitions allow one to explore most of the
parameter space contained in $\chi$, in particular when measuring
$\beta$-spin correlations as function of direction. The full
expression is given  in ref.\cite{JN2013}. The parameter $\xi_2$,
which gives the dependence on spin orientation, is relatively
simple:
\begin{equation}\label{ourthing}
   \xi_2\hat{n}_2^l = A\epsilon^{lmk}\chi_i^{mk},
\end{equation}
where A is the SM $\beta$-asymmetry parameter. In terms of the
generic measurement of $\xi_2$ and thus also for the experiment on
$^{20}$Na described here, one finds that
\begin{equation}\label{ourthing2}
    j^l\epsilon^{lmk}\chi_i^{mk}=\frac{R_\beta^\uparrow +
    R_\beta^\downarrow}{R_\beta^\uparrow - R_\beta^\downarrow}
    \frac{R_\gamma^\uparrow - R_\gamma^\downarrow}{R_\gamma^\uparrow + R_\gamma^\downarrow}.
\end{equation}
 This may suggest that the experimental method just consists of
measuring the  rates in the detectors shown in fig.~\ref{setup}.
However, in particular for a short-lived sample, one would need to
guarantee that conditions during the two polarization periods are
identical.

For completeness we note that in our theory, for any allowed
transition,
\begin{equation}\label{notourthing}
    \xi_3\hat{n}_3=\mp\sqrt{(1-(\alpha Z)^2)(1-\beta^2)}
    \xi_1\hat{n}_1,
\end{equation}
where $\mp$ refers to the case of $\beta^\mp$ decay. Measuring the
$\beta$ polarization is not well possible with high efficiency.
Moreover, $ \xi_3\hat{n}_3 < \xi_1\hat{n}_1$. In this respect
nothing is gained over measuring the $\beta$-emission direction
only.

Of course our theory need not be restricted to $\beta$ decay but
can be evaluated for any weak interaction involving the W boson.
In this respect it is interesting to note that the KLOE
collaboration has measured the lifetime of $K_S$ mesons as
function of the kaon emission direction  with respect to the
dipole anisotropy of the Cosmic Microwave Background \cite{KLOE}.
We find \cite{Vos} that their search for anisotropy is
complementary to $\beta$ decay, however, to gain the maximal
information on $\chi^{\mu\nu}$ a reanalysis of their data would be
required.

In summary, we have identified, in the context of experimental
searches for physics beyond the Standard Model, an important class
of tests for Lorentz symmetry breaking in the weak interaction.
These tests can be made in experiments exploiting the properties
of $\beta$ decay. First experiments have been done. A theoretical
framework that considers modification of the W propagator allows
one to put the various tests in context and relate them to each
other and to non-leptonic weak decays. Part of this theoretical
program has been completed. The theory has a rich structure, for example that there
need not be a single preferred direction for Lorentz invariance
violation. A variety of experiments is required to limit its parameters $\chi^{\mu\nu}$ that, in turn,
 can be related to parameters of the SME model.

This research was supported by the Stichting voor Fundamenteel
Onderzoek der Materie (FOM) under Programmes 104 and 114,
 and FOM projectruimte 08PR2636-1.


\begin{thebibliography}{99}
\bibitem{Kostelecky:2008ts}
  V.~A.~Kostelecky and N.~Russell,
  Rev.\ Mod.\ Phys.\  {\bf 83} (2011) 11, and
arXiv:0801.0287 [hep-ph].

\bibitem{Kostelecky:1988zi}
  V.~A.~Kostelecky and S.~Samuel,
  Phys.\ Rev.\ D {\bf 39} (1989) 683.

\bibitem{Kostelecky:1995qk}
  V.~A.~Kostelecky and R.~Potting,
  Phys.\ Lett.\ B {\bf 381} (1996) 89.
\bibitem{Ellis99} J. R. Ellis, N. E. Mavromatos, and D. V. Nanopoulos, 	arXiv:gr-qc/9909085.
\bibitem{Burgess02}C. P. Burgess, J. M. Cline, E. Filotas, J. Matias, and G. D. Moore, JHEP {\bf 0203} (2002) 043 [hep-ph/0201082].
\bibitem{Gambini99} R. Gambini and J. Pullin, Phys. Rev. D {\bf 59},  (1999) 124021[gr-qc/9809038].
\bibitem{Colladay:1998fq}
  D.~Colladay and V.~A.~Kostelecky,
  Phys.\ Rev.\  D {\bf 58} (1998) 116002.
\bibitem{JN2013} J.P.~Noordmans, H.W.~Wilschut, R.G.E.~Timmermans,
arXiv:1302.2730.
\bibitem{Jackson} J.D.~Jackson, S.B.~Treiman, H.W.~Wyld Jr., Nucl. Phys. {\bf 4} (1957) 206.\\
The SM has also terms with $\vec{J}\cdot\vec{\sigma}$ and $(\vec{\beta}\cdot\vec{\sigma})(\vec{\beta}\cdot\vec{J})$, corresponding to the $N$ and $Q$ correlation coeficients, respectively. These are not discussed here, they are included in~\cite{JN2013}.

\bibitem{Newman:1976sw}
  R.~Newman and S.~Wiesner,
  Phys.\ Rev.\ D {\bf 14} (1976) 1.

\bibitem{Ullman:1978xy}
  J.~D.~Ullman,
  Phys.\ Rev.\ D {\bf 17} (1978) 1750.
\bibitem{Jacob} J.P~Noordmans, H.W.~Wilschut, R.G.E.~Timmermans, in
preparation.
\bibitem{Mueller} S.E.~M\"{u}ller {\it et al.}, in preparation.
\bibitem{neutron}A. Kozela et al., In the proceedings of the Fifth Meeting on
CPT and Lorentz Symmetry, 2010, V.A.~Kostelecký editor:\\
     As a by-product of measuring the $R$ correlation parameter in the decay
     of the neutron, an analysis along the lines of eq.~(\ref{LIVbeta})
     was done leading to limits of order $10^{-2}$. See also~\cite{JN2013} for comments.
\bibitem{Colladay:2001wk}
  D.~Colladay and V.~A.~Kostelecky,
  Phys.\ Lett.\ B {\bf 511} (2001) 209.

\bibitem{Brantjes} N.P.M. Brantjes {\it et al.}, Nucl.\ Instr.\ and Methods A
{\bf 664} (2012) 49.
\bibitem{wilschut} H.W. Wilschut {\it et al.}, Nucl.\ Phys.\ A {\bf 844}
(2010)  143c.
\bibitem{Traykov} E. Traykov {\it et al.} Nucl.\ Instr.\ and Methods A  {\bf 572}
(2007) 580.
\bibitem{Agostinelli:2002hh}
  S.~Agostinelli {\it et al.}  [GEANT4 Collaboration],
  Nucl.\ Instrum.\ Meth.\ A {\bf 506} (2003) 250.

\bibitem{AllisonG4}
J.~Allison {\it et al.},
IEEE Transactions on Nuclear Science 53 No. 1 (2006) 270.

\bibitem{KLOE} F.~Ambrosino {\it et al.}, KLOE Coll., Eur. Phys. J. C  {\bf71} (2011)
1604.
\bibitem{Vos} K. Vos, J.P~Noordmans, H.W.~Wilschut, R.G.E.~Timmermans, in
preparation.
\end{thebibliography}
\end{document}